\documentclass[10pt,journal=jacsat,manuscript=article,layout=twocolumn]{achemso}

\usepackage{chemformula} 
\usepackage[T1]{fontenc} 
\usepackage{caption}
\usepackage{lipsum}
\usepackage{placeins}
\usepackage{xfrac}
\usepackage{siunitx}
\usepackage{booktabs}
\usepackage[table]{xcolor}
\usepackage{array}
\usepackage[font={footnotesize}]{caption}

\let\oldmaketitle\maketitle
\let\maketitle\relax
\author{José P. Carvalho}
\email{jose.carvalho@inano.au.dk}
\author{Anders Bodholt Nielsen}
\author{David L. Goodwin}
\author{Nino Wili}
\author{Niels Chr. Nielsen}
\email{ncn@chem.au.dk}
\affiliation[University] {
Interdisciplinary Nanoscience Center (iNANO) and Department of Chemistry, Aarhus University, Gustav Wieds Vej 14, DK-8000 Aarhus C, Denmark.
}

\title[]
  {Longitudinal Pulsed Dynamic Nuclear Polarization Transfer via Periodic Optimal Control}


\begin{document}

\begin{tocentry}

  \includegraphics[width=\columnwidth]{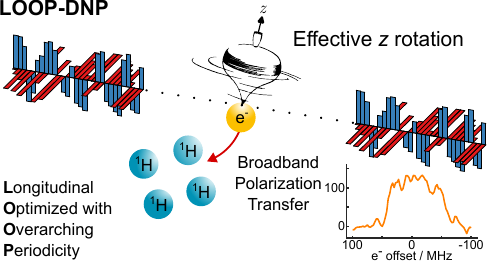}

\end{tocentry}

\twocolumn[
\begin{@twocolumnfalse}
\oldmaketitle
\begin{abstract}
Taking inspiration from NMR spectroscopy, periodic irradiation schemes have recently shown remarkable performance when implemented into pulsed dynamic nuclear polarization (DNP) sequences. This has prompted considerable interest in development of broadband pulsed DNP sequences utilizing such schemes. On this background, most efforts have focused on solid-state NMR like transverse spin-locked pulse sequences whose performance in DNP applications may be compromised by the broadband capabilities of the initial excitation pulse. Leveraging the flexibility and robustness of optimal control theory combined with underlying insights from effective Hamiltonian theory, we present a new family of broadband DNP pulse sequences, termed LOOP (Longitudinally Optimized with Overarching Periodicity), that alleviates the excitation-pulse challenge by accomplishing longitudinal polarization transfer. These sequences define robust single-spin effective $z$ rotations, with impressive compensation towards microwave field inhomogeneity, and are capable of delivering DNP transfer with bandwidths exceeding 100 MHz, while employing a peak microwave field amplitude of only 32 MHz, at an external magnetic field of 0.35 T.
\end{abstract}
\end{@twocolumnfalse}
]
\fontsize{10}{12}\selectfont
In the past two decades, dynamic nuclear polarization (DNP) \cite{OverhauserDNP,Slichter_DNP,Abragam1978} has evolved to become an important element in the toolbox of the Nuclear Magnetic Resonance (NMR) spectroscopist, leveraging the higher polarization of unpaired electrons to enhance nuclear polarization by orders of magnitude. \cite{Becerra_DNP,Corzelius_DNP_review,Corzelius_DNP_review2,leskes_dnp_review} At present, the vast majority of DNP applications rely on continuous-wave (CW) microwave (MW) irradiation, exploiting the Overhauser DNP \cite{OverhauserDNP}, solid-effect \cite{abragam1958nouvelle,jeffries1957polarization}, cross-effect \cite{kessenikh1963proton,hill1967}, and thermal-mixing \cite{provotorov1962magnetic,Borghini1968}  mechanisms. 

\begin{figure}[ht!]
  \centering
  \includegraphics[width=\columnwidth]{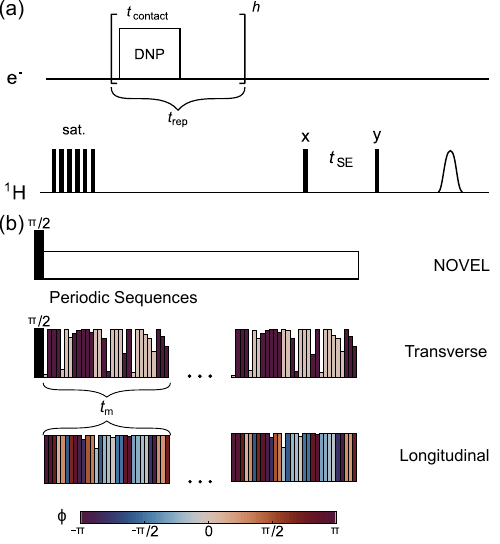}
  \caption{Schematic representation of pulsed DNP sequences used in this work. (a) Pulse sequence scaffold used to measure the DNP enhancement, consisting of $^1$H saturation train, followed by the DNP block where the total buildup time is defined by $T_\text{DNP}=h\cdot t_\text{rep}$, and ending on $^1$H spin polarization readout with a solid-echo sequence. (b) Pulse sequences used for the DNP contact: NOVEL, periodic transverse spin-lock, and periodic  longitudinal schemes.}
  \label{fig:fig1}
\end{figure}

An emerging alternative is pulsed DNP, where polarization transfer is mediated by MW pulse sequences with associated increased control over the spin dynamics \cite{tan2019pulsed}. Various pulsed DNP sequences have been proposed, including various variants of NOVEL (Nuclear-spin orientation Via Electron-spin Locking) \cite{henstra1988nuclear,jain2017off,can2017ramped}, integrated solid-effect  \cite{henstra1988enhanced,can2017frequency}, or adiabatic and stretched solid-effect \cite{tan2020adiabatic,quan2022integrated}. Recently, focus has been devoted to periodic pulse sequences, originally developed with inspiration from solid-state NMR, where irradiation schemes are repeated periodically. Such pulse sequences include time-optimized DNP (TOP DNP) \cite{tan2019time}, PulsePOL \cite{schwartz2018robust}, low-power X-inverse X (XiX) \cite{redrouthu2022efficient}, two-pulse phase modulation (TPPM) \cite{redrouthu2023dynamic}, broadband excitation by amplitude modulation (BEAM) \cite{wili2022designing},  PoLarizAtion Transfer via non-linear Optimization (PLATO) \cite{nielsen2024dynamic}, and constrained Random-Walk Optimized (cRW-OPT) DNP pulse sequences \cite{nielsen2025controlling}  gradually improving the excitation bandwidth. A common trait of these sequences is that the modulation frequency becomes part of the resonance condition, which significantly increases the flexibility to obtain efficient DNP transfer at relatively low MW nutation frequencies or for accomplishing broadband DNP transfers. These sequences may be implemented as illustrated in Figure \ref{fig:fig1}, representing an overall pulse sequence scaffold (Fig. \ref{fig:fig1}a) including a repeated DNP element (Fig. \ref{fig:fig1}b) for e$^-$ $\rightarrow$ $^1$H transfer.

 Aimed at robust DNP in radical systems with large electron paramagnetic resonance (EPR) linewidths, while also coping with notable hardware challenges, such as  MW power limits, inhomogeneity, and phase transients, numerical optimization has become an attractive tool for pulse sequence design \cite{tovsner2009optimal}. Gradient-based optimal control methods \cite{Khaneja2005,OCsolid,OCKuprov} adapted to optimize periodic sequences efficiently for pulsed DNP with the inclusion of phase modulation have expanded the scope of this approach \cite{carvalho2025optimal}. Exploiting a low number of variables in periodic pulse sequences and optimization of effective Hamiltonians, the broadbanded PLATO and cRW-OPT sequences were developed using non-linear optimization methods for an electron-nucleus two-spin model governed by the Hamiltonian \cite{jeschke1996generation,Jeschke_book} 
\begin{equation}
  \mathcal{ \hat H}(t)=\Delta\omega_\text{S} \hat S_z+\omega_I \hat I_z+A\hat S_z\hat I_z+B\hat S_z\hat I_x+\hat{\mathcal H}_\text{MW}(t) .
\end{equation}
Here $\Delta\omega_\text{S}=\omega_\text{e}-\omega_\text{MW}$ represents the MW offset frequency and $\omega_e$ and $\omega_I$  the electron and nuclear spin Larmor frequencies, respectively, all in angular frequencies.  $\hat S_i$ and $\hat I_i$ ($i = x, y, z$) represent Cartesian single-spin operators for the electron and nuclear spins, respectively, and $A$ and $B$ denote the amplitudes for the secular and pseudo-secular hyperfine couplings. $\mathcal{H}_\text{MW}(t)$ corresponds to the MW pulse sequence with irradiation at the MW carrier frequency $\omega_{MW}$. The developed sequences demonstrated broadband DNP with a remarkable agreement between theory, simulations, and experiments, illustrating the efficacy of the underlying design principles. Notably, due to their considerable MW inhomogeneity compensation, the sequences also deliver high transfer efficiencies. A limitation of these sequences, however, is the fact that they are both transverse spin-locked and, therefore, require an initial excitation pulse, as shown in Fig. \ref{fig:fig1}(b), which inevitably may compromise their transfer efficiency at larger offsets, as also shown in Ref. \cite{nielsen2025controlling}. Complementary results are shown in Fig. 3 of the Supporting Information (SI) comparing the DNP transfer efficiency of NOVEL, PLATO, and cRW-OPT1 as a function of the electron spin offset, with an excitation pulse following the electron offset and an always resonant excitation pulse. The decrease in transfer efficiency in the latter is quite noticeable at larger offsets, indicating that the bandwidth of the pulse sequence is indeed limited by the excitation pulse.

In this work, we report new broadband periodic pulse  sequences for DNP polarization transfer without the need for an initial excitation pulse. This is accomplished through the realization of polarization transfer with periodic pulse sequences taking the longitudinal polarization of the unpaired electron spin ($\hat S_z$) to longitudinal nuclear spin polarization ($\hat I_z$). These pulse sequences will henceforth be denoted as LOOP (Longitudinally Optimized with Overarching Periodicity). They are developed using the replicated state-to-state optimal control method proposed in Ref. \cite{carvalho2025optimal} to indirectly optimize the effective Hamiltonian for a series of 24, 25, and 30 phase-modulated pulses, each with 5 ns duration and a peak MW amplitude of 32 MHz (\textit{cf.} Tab. I in the SI). To obtain longitudinal polarization transfer, the LOOP pulses act on the electron spin as an effective $z$ rotation over one basic pulse sequence element. These repeated rotations, in turn, induce a rotation in the Zero-Quantum (ZQ) or Double-Quantum (DQ) subspaces, leading to polarization transfer. The duration of the pulse sequence elements, $\tau_m$, set the modulation frequencies, which for the chosen LOOP sequences corresponded to 8.333, 8.0, and 6.667 MHz, respectively. The choice of modulation time follows the resonance condition
\begin{equation}
\omega_{0I}+k_I \omega_m +\omega^{(S)}_\text{eff}=0,
\end{equation}
where $\omega_{OI}$ corresponds to the nuclear spin Larmor frequency, $k_I$ an integer, and $\omega^{(S)}_\text{eff}$ the electron spin effective field defined by the MW pulse sequence. The choice of modulation time may in this way be used to effectively constrain the required magnitude of the effective field. Considering an external magnetic field of 0.35 T, corresponding to an electron spin resonance frequency in the X-band (9.7 GHz) and to an $^1$H nuclear Larmor frequency of $\omega_{0I}/(2\pi$) = 14.8 MHz, the lowest effective field corresponds to $\omega^{(S)}_\text{eff}/(2\pi)$ values of 1.867, 1.200, and 1.467 MHz for the three considered modulation frequencies, respectively. This range of effective fields is sufficiently large to ensure the polarization transfer can be accurately described by an effective ZQ or DQ Hamiltonian for an anisotropy of the hyperfine coupling tensor of $T/(2\pi)=0.8676$ MHz (corresponding to an electron-nuclear distance of 4.5 \AA, assuming a point-dipole model) \cite{carvalho2025bridging,nielsen2025controlling}, preventing potential destructive interference between the two. At the same time, the effective fields are small enough to avoid excessive sensitivity to  MW inhomogeneity increasing linearly with the effective field. Using these settings and optimizing for sequences with a DNP transfer bandwidth of $\pm$ 50 MHz, we derived a series of LOOP sequences, with LOOP-1$-$5 investigated further in this work (pulse sequences are given in SI). In the optimization, the maximum MW amplitude was limited to 32 MHz, and an MW inhomogeneity profile following a power model \cite{GUPTA201517} was assumed, as detailed in the Materials and Methods section.

\begin{figure*}[ht!]
  \centering
  \includegraphics[width=\textwidth]{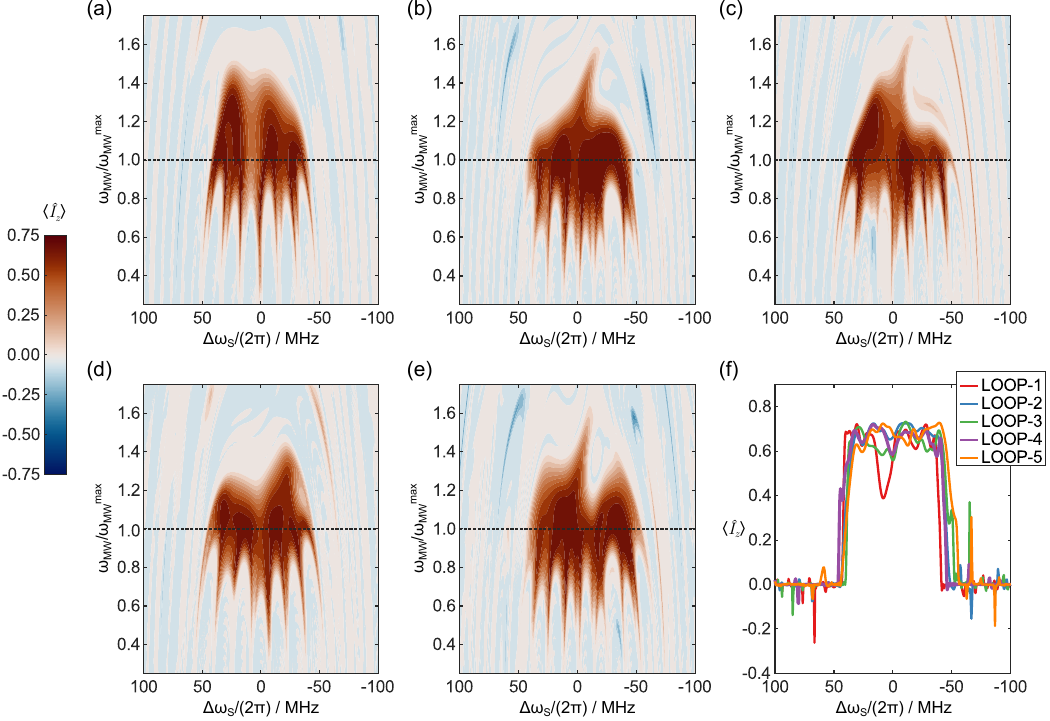}
  \caption{Simulated LOOP DNP transfer profiles as a function of the electron spin offset ($\Delta\omega_\text{S}/(2\pi)$) and deviation from the nominal MW field amplitude ($\omega_{MW}/\omega_{MW}^\text{max}$). The simulations assumed a powder-averaged e$^-$-$^1$H two-spin system with an anisotropy of the hyperfine coupling of $T/(2\pi)$ = 0.8676 MHz (corresponding to an inter-spin distance of $r_\text{eH}$ = 4.5 \AA\,in a point-dipole model) and a static magnetic field corresponding to a proton Larmor frequency of $-14.8$ MHz (approximately 0.35 T).The $t_\text{contact}$ was defined to maximize the integrated DNP transfer bandwidth between $-30$ and 30 MHz. In panels (a)-(e), the transfer profiles are shown as a function of electron spin offset and MW field amplitude. In panel (f), the 1D traces, taken at the nominal MW field strength, $\omega_{MW}^\text{max}/(2\pi)=32$ MHz, are shown.}
  \label{fig:fig2}
\end{figure*}

By numerical simulations, Fig. \ref{fig:fig2} demonstrates the performance of the selected five LOOP DNP pulse sequences for a two-spin e$^-$--$^1$H spin pair using the fidelity function
\begin{equation}
\langle \hat I_z \rangle= \text{Tr}\left[ \hat I_z \,\hat U(T) \hat \rho(0)\hat U^\dagger(T) \right ],
\end{equation} with $\hat\rho(0)=\hat S_z$ and
\begin{equation}
\hat U(T)=\exp_{(\text{O})}\left[-\mathrm{i} \int_0^{T
}\mathcal{\hat H}(t) dt\right],
\end{equation} where $\exp_{(\text{O})}$ denotes the time-ordered exponential operation. As shown in Figs.  \ref{fig:fig2}a-e, all pulse sequences achieve very high DNP transfer efficiency and demonstrate remarkable robustness towards MW offset and inhomogeneity. Upon employing the optimal MW field strength, 
$\omega_{MW}^\text{max}/(2\pi)$ =  32 MHz (\text{cf.} Fig. \ref{fig:fig2}f), all sequences demonstrate a polarization transfer bandwidth approaching 100 MHz. Some slight fissures in transfer efficiency, particularly at MW field strengths deviating from the nominal MW field strength $\omega_{MW}^\text{max}/(2\pi)=32$ MHz, hints at constraints imposed by the modulation time and number of pulses. Further evidence for this interpretation comes from the more pronounced fissures for the LOOP-1 and LOOP-3 sequences, shown in Figs. \ref{fig:fig2}a and c, which have the lowest number of pulses and, therefore, the smallest modulation time.  We note, however, that these small fissures will not have a significant impact on the experimental transfer efficiency, since, experimentally, the transfer profiles are convolved with the EPR lineshape \cite{PLATO_adv}. At higher deviations from the nominal MW field amplitude and large MW offsets, the stability of the effective Hamiltonian starts to become compromised and sensitive to offset, and, therefore, the sequences cannot maintain polarization transfer. At specific combinations of nominal MW field amplitude and MW offset, however, we still observe resonance conditions with the effective field axis also along $z$, leading to peripheral ridges where DNP transfer occurs extending away from the optimal region at which all LOOP sequences operate.

\begin{figure*}[ht!]
  \centering
  \includegraphics[width=\textwidth]{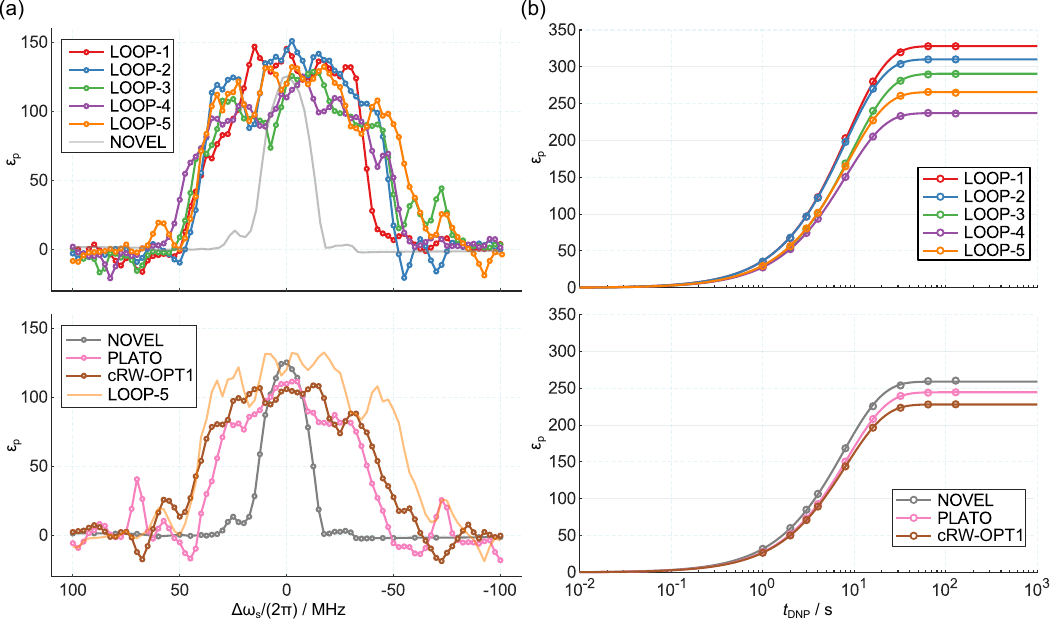}
  \caption{Experimental DNP transfer profiles ($\epsilon_\text{P}$) obtained at X-band on a sample of trityl (OX063) in a water/glycerol mixture at 80K following the pulse sequence represented in Fig. \ref{fig:fig1}, shown above, and the following transverse spin-locked sequences: NOVEL, PLATO, and CRW-1, shown below. (a) DNP transfer profiles as a function of electron spin offset ($\Delta\omega_\text{S}/(2\pi)$) recorded at $t_\text{DNP}=5$ s. To facilitate comparison, faded representations of  the NOVEL and LOOP-5 DNP transfer profiles were shown in the LOOP and transverse spin-locked subplots, respectively. (b) DNP transfer profiles as a function of pumping time $(t_\text{DNP})$ recorded at $\Delta\omega_\text{S}/(2\pi)=0$ MHz. The experimental data was fit to the exponential function: $\epsilon_\text{P}(t_\text{DNP})=\epsilon_\text{max}[1-\exp(-t_\text{DNP}/T_\text{B})]$, where $T_\text{B}$ defines the characteristic buildup time. All fitted buildup times were approximately 8 s (\textit{cf.} Tab. II of the SI). The experimental DNP transfer profiles were all phased so that a positive enhancement occurs at $\Delta\omega_\text{S}/(2\pi)=0$ MHz.}
  \label{fig:fig3}
\end{figure*}

To experimentally validate the five LOOP sequences, pulsed DNP experiments were carried out at X-band on a sample of trityl (OX063) in a water/glycerol solution at 80 K.  Figure \ref{fig:fig3} shows the experimental DNP transfer profiles for PLATO, NOVEL, cRW-OPT1, and five LOOP sequences as a function of electron offset and as a function of pumping time ($t_\text{DNP}$). The offset profiles, all recorded at $t_\text{DNP}=5$ s, show DNP transfer bandwidths of roughly 100 MHz for all LOOP sequences, consistent with simulations and matching cRW-OPT1. We note that some features of experimental features are not directly reproduced in simulations. We attribute this to the simplicity of the model, \textit{i.e.} approximating a large spin system to a two-spin model and neglecting $g$-anisotropy and the response (e.g., phase transients) of the MW resonator.\cite{jegadeesan2025simulation} Nevertheless, the simplicity of the model offers a significant advantage in computational time while also being sufficiently accurate for the purposes of designing effective pulse sequence elements, as demonstrated by a reasonably close match between experimental and simulated offset profiles.  At zero offset, the enhancement factors all exhibit a characteristic buildup time ($T_B$) of roughly 8 s (\textit{cf.} Tab. II of the SI). The LOOP-$1-3$ and $5$ sequences generally achieve higher enhancements than transverse spin-locked sequences. In particular, LOOP-1 leads to roughly a $20\%$ higher enhancement than NOVEL. The higher enhancements for the LOOP sequences can be partially ascribed to the elimination of losses due to the excitation pulse, as well as a better match to the response of the MW resonator (incl. inhomogeneity). Other factors such as multiple-spin effects and relaxation may also contribute; however, a thorough investigation of such effects lies beyond the scope of this study.

In this work we have presented a series of LOOP DNP pulse sequences achieving broadband longitudinal polarization transfer. While their strong performance relative to previous methods by itself merits presentation, this contribution also serves to emphasize the flexibility of pulse sequence design via effective Hamiltonians, in this case illustrated by broadband DNP sequences not  constrained by the performance of an excitation pulse. We expect the LOOP sequences to find immediate applications in low-field DNP while also serving as candidates for bridging to solid-state MAS NMR dipolar recoupling \cite{carvalho2025bridging}. Moreover, as the LOOP elements generate an effective $z$ rotation, they may be readily implemented in DNP echo detection pulse sequences \cite{wili2024observation} and their wider applications for quantum sensing purposes.

\section{Materials and Methods}

All experiments were conducted on a home-built X-band pulsed EPR/DNP spectrometer (based on the design of Doll \textit{et al.} \cite{Doll:2017wb}) equipped with a Keysight M8190A 14 bit/8GSa and Zurich Instruments HDWAG4 16 bit/2.4 GSa arbitrary waveform generators on the MW (electron spins) and RF (nuclear spins) channels, respectively, a 2 kW Applied Systems Engineering 176 X TWT MW amplifier, a SpinCore  GX-5 iSpin NMR console (SpinCore Technologies Inc., Gainesville, FL), a 300 W NanoNord RF amplifier (NanoNord A/S, Aalborg, Denmark), and a Bruker MD4 electron-nuclear double resonance probe (Bruker BioSpin, Rheinstetten, DE) extended with an external tuning and matching circuit. All experiments used a sample of 5 mM trityl (OX063) in a H$_2$O:D$_2$O:Glycerol-d$_8$ solution (1:3:6 by volume) at 80 K. Note that this sample at this temperature is a common system for pulsed-DNP studies. \cite{tritylradicals}

Pulsed DNP experiments were conducted using the  pulse sequence  shown in Fig. \ref{fig:fig1}a using initial saturation of 
$^1$H with a set of $S=11$ pulses of duration 1.32 $\mu$s, separated by $\tau_{\text{sat}}=1$ ms, and a solid-echo sequence  $\pi / 2 - \tau - \pi / 2$  with $\tau$ = 25 $\mu$s for readout. All RF pulses used an  RF field strength of 227.3 kHz, corresponding to a $\pi/2$ pulse time of 1.10 $\mu$s. For the transverse spin-locked sequences, the initial electron spin $(\pi/2)_y$ pulse used an MW field strength of 45.5 MHz, corresponding to a pulse length of 5.5 ns. A 5 s overall pumping time with  $P$ = 2500 and a repetition time of $\tau_{\text{rep}}=2$ ms was used for all experiments, except for the DNP transfer profiles recorded as function of pumping time. Experimental polarization enhancements (denoted $\epsilon_{\text{p}}$) are defined as the ratio between the DNP-enhanced signal intensity and the thermal equilibrium signal intensity (recorded with a repetition time of 120 s, $5\cdot T_{1,n}=112$ s). The specific parameters pertaining to all tested pulse sequences were established via experimental optimization and are shown in Tab. II of the SI. The electronic relaxation time, $T_{1,e}=1.7$ ms, was determined using an inversion recovery sequence, utilizing a frequency-swept MW pulse for the inversion with a linear sweep width of 100 MHz, a total length of 1 $\mu$s, and a quarter-sine amplitude smoothing of 50 ns, followed by spin-echo detection using a $\pi/2$ and $\pi$ pulses with a length of 16 ns and 32 ns and an inter-pulse delay of 400 ns (see Fig. 1a of the SI). The nuclear relaxation time, $T_{1,n}=22.4$ s, was determined by the decay of polarization after DNP employing the NOVEL sequence with $t_\text{DNP}=128$ s (see Fig. 1(b) of the SI).  
$^1$H with a set of $S=11$ pulses of duration 1.32 $\mu$s, separated by $\tau_{\text{sat}}=1$ ms, and a solid-echo sequence  $\pi / 2 - \tau - \pi / 2$  with $\tau$ = 25 $\mu$s for readout. All RF pulses used an  RF field strength of 227.3 kHz, corresponding to a $\pi/2$ pulse time of 1.10 $\mu$s. For the transverse spin-locked sequences, the initial electron spin $(\pi/2)_y$ pulse used an MW field strength of 45.5 MHz, corresponding to a pulse length of 5.5 ns. A 5 s overall pumping time with  $P$ = 2500 and a repetition time of $\tau_{\text{rep}}=2$ ms was used for all experiments, except for the DNP transfer profiles recorded as function of pumping time. Experimental polarization enhancements (denoted $\epsilon_{\text{p}}$) are defined as the ratio between the DNP-enhanced signal intensity and the thermal equilibrium signal intensity (recorded with a repetition time of 120 s, $5\cdot T_{1,n}=112$ s). The specific parameters pertaining to all tested pulse sequences were established via experimental optimization and are shown in Tab. II of the SI. The electronic relaxation time, $T_{1,e}=1.7$ ms, was determined using an inversion recovery sequence, utilizing a frequency-swept MW pulse for the inversion with a linear sweep width of 100 MHz, a total length of 1 $\mu$s, and a quarter-sine amplitude smoothing of 50 ns, followed by spin-echo detection using a $\pi/2$ and $\pi$ pulses with a length of 16 ns and 32 ns and an inter-pulse delay of 400 ns (see Fig. 1a of the SI). The nuclear relaxation time, $T_{1,n}=22.4$ s, was determined by the decay of polarization after DNP employing the NOVEL sequence with $t_\text{DNP}=128$ s (see Fig. 1(b) of the SI).

The LOOP waveforms were developed via optimization of a series of randomly generated sequences with different modulation times. The numerical optimization of all pulse sequences was carried out using the replicated state-to-state procedure described in Ref. \cite{carvalho2025optimal}, employing in-house scripts implemented in C++\cite{ISO/IEC(2020)} using the Eigen3 \cite{eigenweb} library of templates for linear algebra, parallelized at a high level with OpenMP \cite{dagum1998openmp}. All LOOP pulse sequences were optimized for a bandwidth of 90-100 MHz, considering a single e$^-$--$^1$H spin pair with hyperfine couplings of $A/(2\pi)=-0.40$ MHz and $B/(2\pi)=1.00$ MHz, a proton Larmor frequency of $14.8$ MHz, with the MW field inhomogeneity represented by nine MW field strength  scaling factors {0.65, 0.70, 0.75, 0.80, 0.85, 0.90, 0.95, 1.00, 1.05} and corresponding weights {0.079, 0.083, 0.088, 0.094, 0.103, 0.115, 0.135, 0.209, 0.095} defined by a power model \cite{GUPTA201517} and constrained to a maximal MW field of 32 MHz. 

\FloatBarrier
\begin{acknowledgement}
  \fontsize{10}{12}\selectfont
The authors acknowledge advice from Dr. A. Doll (Paul Scherrer Institut, CH), Dr. D. Klose (ETH, Z$\ddot{\text{u}}$rich, CH), and Prof. G. Jeschke (ETH, Z$\ddot{\text{u}}$rich, CH) on building the pulsed X-band EPR/DNP instrumentation. We acknowledge  support from the Novo Nordisk Foundation (NERD grant NNF22OC0076002), the Villum Foundation Synergy programme (grant 50099), and the DeiC National HPC (g.a. DeiC-AU-N5-2024094-H2-2024-35).
\end{acknowledgement}
\bibliography{main-bib}

@book{Jeschke_book,
	author = {Schweiger, Arthur and Jeschke, Gunnar},
	publisher = {Oxford university press},
	title = {Principles of pulse electron paramagnetic resonance},
	year = {2001}}

@article{jeschke1996generation,
  title={Generation and transfer of coherence in electron-nuclear spin systems by non-ideal microwave pulses},
  author={Jeschke, Gunnar},
  journal={Molecular Physics},
  volume={88},
  number={2},
  pages={355--383},
  year={1996},
  publisher={Taylor \& Francis}
}

@article{Slichter_DNP,
	author = {Carver, T. R. and Slichter, C. P.},
	journal = {Phys. Rev.},
	month = {Oct},
	pages = {212--213},
	title = {Polarization of Nuclear Spins in Metals},
	volume = {92},
	year = {1953}}

@article{can2017frequency,
  title={Frequency-Swept Integrated Solid Effect},
  author={Can, Thach V and Weber, Ralph T and Walish, Joseph J and Swager, Timothy M and Griffin, Robert G},
  journal={Angewandte Chemie},
  volume={129},
  number={24},
  pages={6848--6852},
  year={2017},
  publisher={Wiley Online Library}
}

@article{schwartz2018robust,
  title={Robust optical polarization of nuclear spin baths using {H}amiltonian engineering of nitrogen-vacancy center quantum dynamics},
  author={Schwartz, Ilai and Scheuer, Jochen and Tratzmiller, Benedikt and M{\"u}ller, Samuel and Chen, Qiong and Dhand, Ish and Wang, Zhen-Yu and M{\"u}ller, Christoph and Naydenov, Boris and Jelezko, Fedor and others},
  journal={Science advances},
  volume={4},
  number={8},
  pages={eaat8978},
  year={2018},
  publisher={American Association for the Advancement of Science}
}

@article{redrouthu2022efficient,
  title={Efficient pulsed dynamic nuclear polarization with the {X}-inverse-{X} sequence},
  author={Redrouthu, Venkata SubbaRao and Mathies, Guinevere},
  journal={Journal of the American Chemical Society},
  volume={144},
  number={4},
  pages={1513--1516},
  year={2022},
  publisher={ACS Publications}
}

@article{redrouthu2023dynamic,
    author = {Redrouthu, Venkata SubbaRao and Vinod-Kumar, Sanjay and Mathies, Guinevere},
    title = "{Dynamic nuclear polarization by two-pulse phase modulation}",
    journal = {The Journal of Chemical Physics},
    volume = {159},
    number = {1},
    pages = {014201},
    year = {2023},
    month = {07},
}

@article{carvalho2025optimal,
  title={Optimal control design strategies for pulsed dynamic nuclear polarization},
  author={Carvalho, Jos{\'e} P and Goodwin, David L and Wili, Nino and Nielsen, Anders Bodholt and Nielsen, Niels Chr},
  journal={The Journal of Chemical Physics},
  volume={162},
  number={5},
  year={2025},
  publisher={AIP Publishing}
}

@article{GUPTA201517,
	author = {Rupal Gupta and Guangjin Hou and Tatyana Polenova and Alexander J. Vega},
	journal = {Solid State Nuclear Magnetic Resonance},
	pages = {17-26},
	title = {RF inhomogeneity and how it controls CPMAS},
	volume = {72},
	year = {2015}}

@article{dagum1998openmp,
  title={{OpenMP}: an industry standard {API} for shared-memory programming},
  author={Dagum, Leonardo and Menon, Ramesh},
  journal={{IEEE} computational science and engineering},
  volume={5},
  number={1},
  pages={46--55},
  year={1998},
  publisher={{IEEE}}
}

@MISC{eigenweb,
  author = {Ga\"{e}l Guennebaud and Beno\^{i}t Jacob and others},
  title = {Eigen v3},
  howpublished = {http://eigen.tuxfamily.org},
  year = {2010}
 }

@article{Doll:2017wb,
	address = {Laboratory of Physical Chemistry, ETH Zurich, Vladimir-Prelog-Weg 2, CH-8093 Zurich, Switzerland.; Laboratory of Physical Chemistry, ETH Zurich, Vladimir-Prelog-Weg 2, CH-8093 Zurich, Switzerland. Electronic address: gjeschke@ethz.ch.},
	author = {Doll, Andrin and Jeschke, Gunnar},
	copyright = {Copyright {\copyright}2017 Elsevier Inc. All rights reserved.},
	crdt = {2017/06/06 06:00},
	date = {2017 Jul},
	doi = {10.1016/j.jmr.2017.01.004},
	edat = {2017/06/06 06:00},
	issn = {1096-0856 (Electronic); 1090-7807 (Linking)},
	jid = {9707935},
	journal = {J Magn Reson},
	jt = {Journal of magnetic resonance (San Diego, Calif. : 1997)},
	lid = {S1090-7807(17)30004-6 {$[$}pii{$]$}; 10.1016/j.jmr.2017.01.004 {$[$}doi{$]$}},
	lr = {20191120},
	mhda = {2017/06/06 06:01},
	month = {Jul},
	oto = {NOTNLM},
	own = {NLM},
	pages = {46--62},
	phst = {2016/11/30 00:00 {$[$}received{$]$}; 2016/12/31 00:00 {$[$}revised{$]$}; 2017/01/03 00:00 {$[$}accepted{$]$}; 2017/06/06 06:00 {$[$}entrez{$]$}; 2017/06/06 06:00 {$[$}pubmed{$]$}; 2017/06/06 06:01 {$[$}medline{$]$}},
	pii = {S1090-7807(17)30004-6},
	pl = {United States},
	pmid = {28579102},
	pst = {ppublish},
	pt = {Journal Article; Research Support, Non-U.S. Gov't},
	status = {PubMed-not-MEDLINE},
	title = {Wideband frequency-swept excitation in pulsed EPR spectroscopy.},
	volume = {280},
	year = {2017}}

@Article{nielsen2024dynamic,
author ="Nielsen, A. B. and Carvalho, J. P. A. and Goodwin, D. L. and Wili, N. and Nielsen, N. C.",
title  ="Dynamic nuclear polarization pulse sequence engineering using single-spin vector effective Hamiltonians",
journal  ="Phys. Chem. Chem. Phys.",
year  ="2024",
volume  ="26",
issue  ="44",
pages  ="28208-28219",
publisher  ="The Royal Society of Chemistry",
}

@article{wili2022designing,
  title={Designing broadband pulsed dynamic nuclear polarization sequences in static solids},
  author={Wili, Nino and Nielsen, Anders Bodholt and V{\"o}lker, Laura Alicia and Schreder, Lukas and Nielsen, Niels Chr and Jeschke, Gunnar and Tan, Kong Ooi},
  journal={Science Advances},
  volume={8},
  number={28},
  pages={eabq0536},
  year={2022},
  publisher={American Association for the Advancement of Science}
}

@article{henstra1988enhanced,
  title={Enhanced dynamic nuclear polarization by the integrated solid effect},
  author={Henstra, A and Dirksen, P and Wenckebach, W Th},
  journal={Physics Letters A},
  volume={134},
  number={2},
  pages={134--136},
  year={1988},
  publisher={Elsevier}
}

@article{henstra1988nuclear,
  title={Nuclear spin orientation via electron spin locking ({NOVEL})},
  author={Henstra, A and Dirksen, P and Schmidt, J and Wenckebach, W Th},
  journal={Journal of Magnetic Resonance (1969)},
  volume={77},
  number={2},
  pages={389--393},
  year={1988},
  publisher={Elsevier}
}

@article{tan2019time,
  title={Time-optimized pulsed dynamic nuclear polarization},
  author={Tan, Kong Ooi and Yang, Chen and Weber, Ralph T and Mathies, Guinevere and Griffin, Robert G},
  journal={Science advances},
  volume={5},
  number={1},
  pages={eaav6909},
  year={2019},
  publisher={American Association for the Advancement of Science}
}

@article{quan2022integrated,
  title={Integrated, stretched, and adiabatic solid effects},
  author={Quan, Yifan and Steiner, Jakob and Ouyang, Yifu and Tan, Kong Ooi and Wenckebach, W Thomas and Hautle, Patrick and Griffin, Robert G},
  journal={The journal of physical chemistry letters},
  volume={13},
  number={25},
  pages={5751--5757},
  year={2022},
  publisher={ACS Publications}
}

@article{tan2020adiabatic,
  title={Adiabatic solid effect},
  author={Tan, Kong Ooi and Weber, Ralph T and Can, Thach V and Griffin, Robert G},
  journal={The journal of physical chemistry letters},
  volume={11},
  number={9},
  pages={3416--3421},
  year={2020},
  publisher={ACS Publications}
}

@article{can2017ramped,
    author = {Can, T. V. and Weber, R. T. and Walish, J. J. and Swager, T. M. and Griffin, R. G.},
    title = "{Ramped-amplitude {NOVEL}}",
    journal = {The Journal of Chemical Physics},
    volume = {146},
    number = {15},
    pages = {154204},
    year = {2017},
    month = {04},
    issn = {0021-9606}}

@article{jain2017off,
    author = {Jain, Sheetal K. and Mathies, Guinevere and Griffin, Robert G.},
    title = "{Off-resonance {NOVEL}}",
    journal = {The Journal of Chemical Physics},
    volume = {147},
    number = {16},
    pages = {164201},
    year = {2017},
    month = {10},
    issn = {0021-9606}}

@article{Becerra_DNP,
	author = {Becerra, Lino R. and Gerfen, Gary J. and Temkin, Richard J. and Singel, David J. and Griffin, Robert G.},
	journal = {Phys. Rev. Lett.},
	month = {Nov},
	pages = {3561--3564},
	title = {Dynamic nuclear polarization with a cyclotron resonance maser at 5 T},
	volume = {71},
	year = {1993}}

@article{Corzelius_DNP_review,
	author = {Aany Sofia {Lilly Thankamony} and Johannes J. Wittmann and Monu Kaushik and Bj{\"o}rn Corzilius},
	journal = {Progress in Nuclear Magnetic Resonance Spectroscopy},
	pages = {120-195},
	title = {Dynamic nuclear polarization for sensitivity enhancement in modern solid-state NMR},
	volume = {102-103},
	year = {2017}}

@article{Corzelius_DNP_review2,
author = {Biedenbänder, Thomas and Aladin, Victoria and Saeidpour, Siavash and Corzilius, Bj{\"o}rn},
title = {Dynamic Nuclear Polarization for Sensitivity Enhancement in Biomolecular Solid-State  {NMR}},
journal = {Chemical Reviews},
volume = {122},
number = {10},
pages = {9738-9794},
year = {2022},
}

@article{leskes_dnp_review,
  title={Dynamic nuclear polarization solid-state {NMR} spectroscopy for materials research},
  author={Moroz, Ilia B and Leskes, Michal},
  journal={Annual Review of Materials Research},
  volume={52},
  number={1},
  pages={25--55},
  year={2022},
  publisher={Annual Reviews}
}

@article{PLATO_adv,
	author = {Nielsen, A. B. and Carvalho, J. P. A. and Goodwin, D. L. and Wili, N. and Nielsen, N. C.},
	journal = {Phys. Chem. Chem. Phys.},
	pages = {28208-28219},
	title = {Dynamic nuclear polarization pulse sequence engineering using single-spin vector effective Hamiltonians},
	volume = {26},
	year = {2024}}

@article{Abragam1978,
year = {1978},
month = {mar},
publisher = {},
volume = {41},
number = {3},
pages = {395},
author = {A Abragam and M Goldman},
title = {Principles of dynamic nuclear polarisation},
journal = {Reports on Progress in Physics},
}

@article{tritylradicals,
author = {Mathies, Guinevere and Jain, Sheetal and Reese, Marcel and Griffin, Robert G.},
title = {Pulsed Dynamic Nuclear Polarization with Trityl Radicals},
journal = {The Journal of Physical Chemistry Letters},
volume = {7},
number = {1},
pages = {111-116},
year = {2016}}

@article{hill1967,
  title = {New Effect in Dynamic Polarization},
  author = {Hwang, Chester F. and Hill, D. A.},
  journal = {Phys. Rev. Lett.},
  volume = {18},
  issue = {4},
  pages = {110--112},
  numpages = {0},
  year = {1967},
  month = {Jan},
  publisher = {American Physical Society},
}

@article{provotorov1962magnetic,
  title={Magnetic resonance saturation in crystals},
  author={Provotorov, BN},
  journal={Soviet Physics Jetp-Ussr},
  volume={14},
  number={5},
  pages={1126--1131},
  year={1962},
  publisher={AMER INST PHYSICS 1305 WALT WHITMAN RD, STE 300, MELVILLE, NY 11747-4501 USA}
}

@article{Borghini1968,
  title = {Spin-Temperature Model of Nuclear Dynamic Polarization Using Free Radicals},
  author = {Borghini, M.},
  journal = {Phys. Rev. Lett.},
  volume = {20},
  issue = {9},
  pages = {419--421},
  numpages = {0},
  year = {1968},
  month = {Feb},
  publisher = {American Physical Society},
}

@article{tan2019pulsed,
  title={Pulsed dynamic nuclear polarization},
  author={Tan, Kong Ooi and Jawla, Sudheer and Temkin, Richard J and Griffin, Robert G},
  journal={Handbook of High Field Dynamic Nuclear Polarization},
  pages={71--86},
  year={2019},
  publisher={John Wiley \& Sons}
}

@article{kessenikh1963proton,
  title={Proton polarization in irradiated polyethylenes},
  author={Kessenikh, AV and Lushchikov, VI and Manenkov, AA and Taran, Yu V},
  journal={Soviet Phys.-Solid State (English Transl.)},
  volume={5},
  year={1963},
  publisher={Lebedev Inst. of Physics, Moscow; Karpov Research Inst. for Physics and~…}
}

@article{abragam1958nouvelle,
  title={Une nouvelle m{\'e}thode de polarisation dynamique des noyaux atomiques dans les solides},
  author={Abragam, Anatole and Proctor, Warren G},
  journal={Comp. Rend. Acad. Sci},
  volume={246},
  pages={2253--2256},
  year={1958}
}

@article{jeffries1957polarization,
  title={Polarization of nuclei by resonance saturation in paramagnetic crystals},
  author={Jeffries, CD},
  journal={Physical Review},
  volume={106},
  number={1},
  pages={164},
  year={1957},
  publisher={APS}
}

@article{OverhauserDNP,
	author = {Overhauser, Albert W.},
	journal = {Phys. Rev.},
	month = {Oct},
	pages = {411--415},
	title = {Polarization of Nuclei in Metals},
	volume = {92},
	year = {1953}}

@article{carvalho2025bridging,
  title={Bridging Dynamic Nuclear Polarization and Solid-State NMR Dipolar Recoupling: From Static Single Crystal to Spinning Powders},
  author={Carvalho, Jos{\'e} P and Nielsen, Anders Bodholt and Baligacs, Eniko and Wili, Nino and Nielsen, Niels Chr},
  journal={The Journal of Physical Chemistry Letters},
  volume={16},
  number={17},
  pages={4363--4371},
  year={2025},
  publisher={ACS Publications}
}

@article{wili2024observation,
  title={Observation of dynamic nuclear polarization echoes},
  author={Wili, Nino and Nielsen, Anders B and Carvalho, Jos{\'e} P and Nielsen, Niels Chr},
  journal={Science Advances},
  volume={10},
  number={42},
  pages={eadr2420},
  year={2024},
  publisher={American Association for the Advancement of Science}
}

@article{nielsen2025controlling,
  title={Controlling effective Hamiltonians: Broadband pulsed dynamic nuclear polarization by constrained random walk and non-linear optimization},
  author={Nielsen, Anders B and Carvalho, Jos{\'e} P and Wili, Nino and Jensen, Filip V and Goodwin, David L and Untidt, Thomas S and To{\v{s}}ner, Zden{\v{e}}k and Nielsen, Niels Chr},
  journal={The Journal of Chemical Physics},
  volume={163},
  number={14},
  year={2025},
  publisher={AIP Publishing}
}

@article{jegadeesan2025simulation,
  title={Simulation of pulsed dynamic nuclear polarization in the steady state},
  author={Jegadeesan, Shebha Anandhi and Zhao, Yujie and Smith, Graham M and Kuprov, Ilya and Mathies, Guinevere},
  journal={arXiv preprint arXiv:2505.24444},
  year={2025}
}

@article{tovsner2009optimal,
  title={Optimal control in {NMR} spectroscopy: Numerical implementation in {SIMPSON}},
  author={To{\v{s}}ner, Zden{\v{e}}k and Vosegaard, Thomas and Kehlet, Cindie and Khaneja, Navin and Glaser, Steffen J and Nielsen, Niels Chr},
  journal={Journal of Magnetic Resonance},
  volume={197},
  number={2},
  pages={120--134},
  year={2009},
  publisher={Elsevier}
}

@article{Khaneja2005,
	doi     = {10.1016/j.jmr.2004.11.004},
	year    = 2005,
	volume  = {172},
	pages   = {296--305},
	author  = {Navin Khaneja and Timo Reiss and Cindie Kehlet and Thomas Schulte-Herbr{\"{u}}ggen and Steffen J. Glaser},
	title   = {Optimal control of coupled spin dynamics: design of {NMR} pulse sequences by gradient ascent algorithms},
	journal = {J. Magn. Reson.}}

@article{OCsolid,
	author = {Kehlet, Cindie T. and Sivertsen, Astrid C. and Bjerring, Morten and Reiss, Timo O. and Khaneja, Navin and Glaser, Steffen J. and Nielsen, Niels Chr.},
	journal = {Journal of the American Chemical Society},
	number = {33},
	pages = {10202-10203},
	title = {Improving Solid-State NMR Dipolar Recoupling by Optimal Control},
	volume = {126},
	year = {2004}}

@article{OCKuprov,
	author = {Goodwin,D. L. and Kuprov,Ilya},
	journal = {The Journal of Chemical Physics},
	number = {20},
	pages = {204107},
	title = {Modified Newton-Raphson GRAPE methods for optimal control of spin systems},
	volume = {144},
	year = {2016}}

\clearpage

\onecolumn
\section{Supporting Information for: Longitudinal Pulsed Dynamic Nuclear Polarization Transfer via Periodic Optimal Control}
\subsection{LOOP Waveforms}
\renewcommand{\thefigure}{S\arabic{figure}}
\renewcommand{\thetable}{S\arabic{table}}
\setcounter{figure}{0}
\setcounter{table}{0}

\begin{table}[h!]
\scriptsize
\setlength\extrarowheight{-5pt}
\color{black}
\rowcolors{2}{blue!10}{white}
\begin{tabular}{
c
|S[table-format=2.3] S[table-format=1.3]
|S[table-format=2.3] S[table-format=1.3]
|S[table-format=2.3] S[table-format=1.3]
|S[table-format=2.3] S[table-format=1.3]
|S[table-format=2.3] S[table-format=1.3]
}
\toprule
 \multicolumn{1}{c|}{}       & \multicolumn{2}{c|}{LOOP-1}        & \multicolumn{2}{c|}{LOOP-2} &\multicolumn{2}{c|}{LOOP-3}   & \multicolumn{2}{c|}{LOOP-4}    & \multicolumn{2}{c}{LOOP-5}       \\
Pulse $i$ & \ensuremath{[\sfrac{\omega_\text{MW}}{2\pi}]_i} &  $\phi_i$ & \ensuremath{[\sfrac{\omega_\text{MW}}{2\pi}]_i} & $\phi_i$ & \ensuremath{[\sfrac{\omega_\text{MW}}{2\pi}]_i} & $\phi_i$ & \ensuremath{[\sfrac{\omega_\text{MW}}{2\pi}]_i} & $\phi_i$ & \ensuremath{[\sfrac{\omega_\text{MW}}{2\pi}]_i} & $\phi_i$ \\
\midrule
1     & 30.594 & -2.026 & 32.000 & 0.334  & 32.000 & 2.346  & 32.000 & -1.635 & 32.000 & -0.042 \\
2     & 23.002 & 0.263  & 32.000 & -0.227 & 28.366 & 1.646  & 31.853 & -0.505 & 32.000 & -0.190 \\
3     & 32.000 & 0.484  & 32.000 & -0.708 & 15.323 & 1.019  & 32.000 & 1.526  & 31.997 & -0.786 \\
4     & 32.000 & -0.661 & 31.999 & -2.443 & 32.000 & 2.503  & 32.000 & 1.999  & 32.000 & -2.792 \\
5     & 28.827 & -0.947 & 32.000 & -2.794 & 31.856 & 2.562  & 32.000 & 2.477  & 32.000 & -2.289 \\
6     & 24.079 & 0.759  & 31.502 & 0.161  & 32.000 & -0.099 & 32.000 & 1.959  & 31.909 & 1.215  \\
7     & 27.672 & 1.304  & 29.055 & -0.885 & 32.000 & -1.023 & 25.728 & 0.546  & 31.957 & -1.042 \\
8     & 23.291 & 1.330  & 32.000 & -0.037 & 29.388 & -1.554 & 20.017 & -0.491 & 32.000 & -0.368 \\
9     & 25.573 & 0.515  & 32.000 & 0.654  & 26.190 & -1.891 & 23.847 & -0.885 & 32.000 & 0.212  \\
10    & 32.000 & -0.906 & 25.720 & -0.359 & 32.000 & -1.088 & 32.000 & 0.254  & 29.161 & 0.626  \\
11    & 31.336 & -0.633 & 32.000 & -1.698 & 28.778 & -0.114 & 32.000 & 1.078  & 32.000 & -1.666 \\
12    & 26.973 & 0.656  & 32.000 & -2.084 & 32.000 & 0.511  & 32.000 & 0.682  & 32.000 & -2.077 \\
13    & 14.547 & 1.040  & 20.204 & 1.313  & 23.773 & 0.064  & 29.382 & -0.324 & 22.890 & 2.537  \\
14    & 30.535 & -2.507 & 28.519 & 1.720  & 20.765 & -1.964 & 23.126 & -0.627 & 32.000 & 1.223  \\
15    & 32.000 & -2.480 & 32.000 & 2.589  & 32.000 & -1.919 & 24.934 & 1.337  & 32.000 & 2.421  \\
16    & 22.776 & -2.935 & 32.000 & 3.083  & 31.959 & -0.963 & 31.959 & 2.406  & 32.000 & 2.568  \\
17    & 23.008 & 3.008  & 26.982 & 2.529  & 32.000 & -1.079 & 31.984 & 2.568  & 30.264 & 2.980  \\
18    & 22.514 & -2.089 & 32.000 & 2.119  & 32.000 & -1.890 & 31.975 & -2.833 & 31.988 & 1.556  \\
19    & 32.000 & -0.899 & 32.000 & 1.242  & 23.824 & 2.897  & 29.889 & 1.568  & 31.993 & 1.562  \\
20    & 32.000 & -0.884 & 31.919 & -0.841 & 16.521 & 1.218  & 31.977 & 1.030  & 31.964 & -0.556 \\
21    & 31.505 & -2.497 & 27.472 & 0.362  & 29.365 & 1.297  & 29.457 & -0.499 & 31.171 & -0.284 \\
22    & 26.951 & 2.878  & 32.000 & 2.053  & 32.000 & 2.182  & 31.994 & 0.389  & 32.000 & 2.153  \\
23    & 26.704 & -2.424 & 32.000 & 2.469  & 32.000 & -2.615 & 31.532 & 1.464  & 32.000 & 2.303  \\
24    & 28.793 & -2.604 & 32.000 & 2.646  & 32.000 & -2.309 & 29.007 & 1.449  & 32.000 & 2.528  \\
25    &      &        & 29.841 & 2.509  &      &        & 31.997 & 0.818  & 32.000 & 2.723  \\
26    &      &        & 29.205 & 0.988  &      &        &      &        & 32.000 & 0.654  \\
27    &      &        & 32.000 & 0.386  &      &        &      &        & 32.000 & 0.169  \\
28    &      &        & 30.076 & -2.418 &      &        &      &        & 31.956 & -2.316 \\
29    &      &        & 32.000 & -2.003 &      &        &      &        & 31.996 & -2.415 \\
30    &      &        & 31.994 & -0.872 &      &        &      &        & 31.998 & -0.863  \\
\bottomrule
\end{tabular}
\caption{The five LOOP DNP pulse sequences described in the main text with MW amplitudes $(\omega_{MW}/(2\pi))_i$ in MHz and phases $\phi_i$ in radians wrapped to the $[-\pi,\pi[$ interval for all pulses of duration 5 ns.}
\end{table}

\subsection{Pulse sequence parameters}

\begin{table}[h!]
\scriptsize
\setlength\extrarowheight{-5pt}
\color{black}
\rowcolors{2}{blue!10}{white}
\begin{tabular}{
c
|S[table-format=3.0]
|S[table-format=2.1]
|S[table-format=3.1]
|S[table-format=1.1]
}
\toprule
Pulse sequence & \ensuremath{t_\text{contact}~(\si{ns})} & \ensuremath{\omega^\text{max}_\text{MW}/(2\pi)~(\si{MHz})} & \ensuremath{\epsilon_\text{max}} & \ensuremath{T_\text{B}~(\si{s})} \\
\midrule
NOVEL     & 8000 & 13.0 & 259 & 7.6\\
PLATO     & 840  & 29.7 & 245 & 8.4\\
cRW-OPT1  & 750  & 35.0 & 228 & 8.0\\
LOOP-1    & 840  & 32.8 & 328 & 8.4\\
LOOP-2    & 900  & 32.0 & 310 & 7.9\\
LOOP-3    & 840  & 25.1 & 291 & 9.3\\
LOOP-4    & 1000 & 26.7 & 237 & 8.0\\
LOOP-5    & 900  & 27.4 & 266 & 8.3\\
\bottomrule
\end{tabular}
\caption{Experimentally optimized parameters for the pulse sequences examined in this work: contact time $t_\text{contact}$, DNP contact maximal MW field strength (note that NOVEL should have a nominal value of 14.8 MHz at this external magnetic field, while for all other sequences $\omega^\text{max}_\text{MW}=32$ MHz), maximal enhancements $\epsilon_\text{max}$, and characteristic buildup times $T_\text{B}$. The enhancement factors and pulse sequence parameters were all obtained at $\Delta\omega_\text{S}/(2\pi)=0$ MHz. See the main text for experimental details.}
\end{table}

\clearpage

\subsection{$T_1$ relaxation times}

\begin{figure*}[h!]
  \centering
  \includegraphics[width=\textwidth]{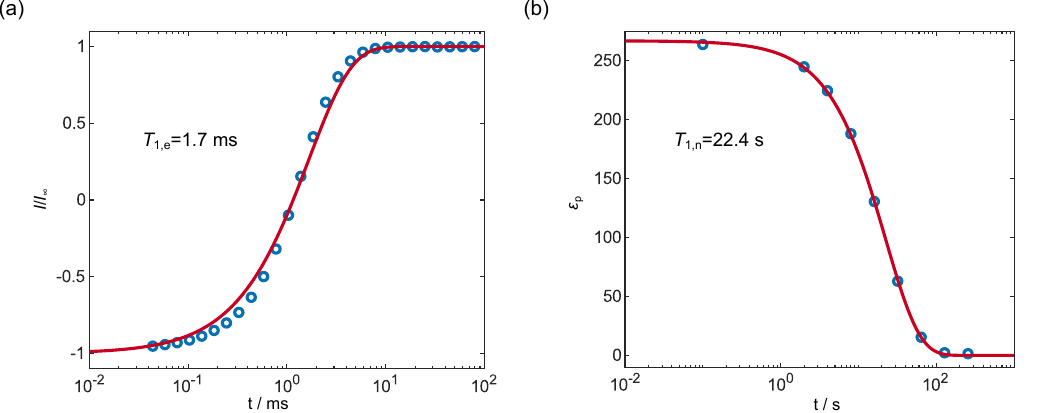}
  \caption{$T_1$ relaxation times. (a) Inversion recovery data for electronic spins. $T_{1,e}$ determined via an exponential fit:$I/I_\infty(t)=[1-2\exp(-t/T_{1,e})]$. (b) Nuclear polarization after DNP. $T_{1,n}$ determined via an exponential fit: $\epsilon_\text{p}(t)=\epsilon_\text{max}\exp(-t/T_{1,n})]$. See the main text for experimental details.}
\end{figure*}

\subsection{Effect of the flip pulse}

\begin{figure*}[h!]
  \centering
  \includegraphics[width=\textwidth]{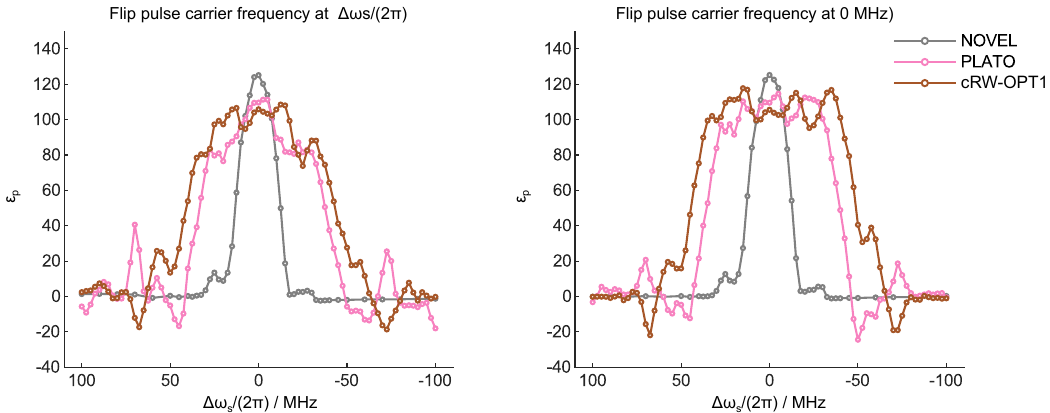}
  \caption{Experimental DNP transfer profiles ($\epsilon_\text{P}$) for NOVEL, PLATO, and cRW-OPT1 pulse sequences as a function of electron spin offset ($\Delta\omega_\text{S}/(2\pi)$) recorded at $t_\text{DNP}=5$ s, obtained at X-band on a sample of trityl (OX063) in a water/glycerol mixture at 80 K, following the pulse sequence represented in Fig. \ref{fig:fig1} of the main text. In the left panel, the initial electron-spin $\pi/2$ pulse follows the offset of the DNP pulse sequence element, whereas in the right panel, it is on-resonance throughout the offset scan of the DNP sequence.}
\end{figure*}

\end{document}